\documentclass[fleqn,usenatbib]{mnras}

\usepackage{newtxtext,newtxmath}
\usepackage[T1]{fontenc}

\DeclareRobustCommand{\VAN}[3]{#2}
\let\VANthebibliography\thebibliography
\def\thebibliography{\DeclareRobustCommand{\VAN}[3]{##3}\VANthebibliography}

\usepackage{graphicx}	% Including figure files
\usepackage{amsmath}	% Advanced maths commands
\usepackage{bm}
\usepackage{here}
\usepackage{breqn}
\usepackage{hyperref}

\title[Coronal Properties of Low-mass Pop III Stars]{Coronal Properties of Low-mass Population III Stars and the Radiative Feedback in the Early Universe}

\author[Washinoue \& Suzuki ]{
Haruka Washinoue $^{1}$\thanks{E-mail: washi@ea.c.u-tokyo.ac.jp}
and Takeru K. Suzuki$^{1}$
\\
$^{1}$ School of Arts and Sciences, University of Tokyo, 3-8-1, Komaba, Meguro, Tokyo 153-8902, Japan\\
}

\date{Accepted XXX. Received YYY; in original form ZZZ}

\pubyear{2015}

\begin{document}

\label{firstpage}
\pagerange{\pageref{firstpage}--\pageref{lastpage}}
\maketitle

\begin{abstract}
We systematically investigated the heating of coronal loops on metal-free stars with various stellar masses and magnetic fields by magnetohydrodynamic simulations. It is found that the coronal property is dependent on the coronal magnetic field strength $B_{\rm c}$ because it affects the difference of the nonlinearity of the Alfv\'{e}nic waves. Weaker $B_{\rm c}$ leads to cooler and less dense coronae because most of the input waves dissipate in the lower atmosphere on account of the larger nonlinearity. Accordingly EUV and X-ray luminosities also correlate with $B_{\rm c}$, while they are emitted in a wide range of the field strength. Finally we extend our results to evaluating the contribution from low-mass Population III coronae to the cosmic reionization. Within the limited range of our parameters on magnetic fields and loop lengths, the EUV and X-ray radiations give a weak impact on the ionization and heating of the gas at high redshifts. However, there still remains a possibility of the contribution to the reionization from energetic flares involving long magnetic loops. 
\end{abstract}

\begin{keywords}
(magnetohydrodynamics) MHD -- cosmology: dark ages, reionization, first stars -- stars: coronae -- stars: low-mass 
\end{keywords}

%%%%%%%%%%%%%%%%%%%%%%%%%%%%%%%%%%%%%%%%%%%%%%%%%%

%%%%%%%%%%%%%%%%% BODY OF PAPER %%%%%%%%%%%%%%%%%%

\section{Introduction}

Radiation from Population III (Pop III) stars are believed to play an important role in the cosmic reionization \citep{LoBa2001, WyLo2003, So_2004, Wi_2008}. 
When we discuss their radiative feedback, the stellar mass is a critical parameter that determines the radiation energy and spectrum. However, their initial mass function (IMF) has not been well understood yet.

Although the mass of Pop III stars was considered to be weighted on the more massive side with $\gtrsim 100M_{\odot}$ \citep{Ab2002, OmPa2003,Yo_2006,OsNo2007}, it has been lately suggested to have a wide range of stellar masses (e.g., \citealp{Hi_2014}). In addition, recent numerical studies revealed that primordial protostellar disks often fragment into multiple small pieces by the gravitational instability \citep{Cl_2011, Gr2011, MaDo2013, St_2016, Su2019, ChYo2020}, while the fate of these fragments are still under debate; on one hand, a fraction of them seem that they migrate inward by the efficient transport of the angular momentum and finally infall to the central protostar. On the other hand, other small fragments ejected out of the halo via many-body gravitational interactions and left as isolated low-mass stars. 
The formation and the existence of low-mass Pop III stars have not been concluded because the sufficient numerical resolutions are not achieved and there is no observational evidence.

There is also an indirect approach to obtain the constraints on the existence and abundance of low-mass Pop III stars through the search for them in the current universe \citep{Ha_2015, Is_2016, Ma_2019}. These studies indicate that non-detection of low-mass Pop III stars so far means that the formation of such stars were highly suppressed in the early universe or they have been polluted by heavy elements during their lifetime.
The effect of the metal pollution by the interstellar medium has been one of the important topics for the observation of Pop III survivors; some studies suggest that the metal accretion can be efficient \citep{Ko_2015, Sh_2017}, while others claim that the stellar winds may prevent it except for large interstellar objects \citep{Ta_2018} and the effect is almost negligible \citep{Ta_2017,Su2018}.

In addition to the uncertainty of their existence and formation processes, there is little focus on the property of low-mass Pop III stars themselves. In particular the radiative properties are greatly different from those of the massive Pop III stars. 

 According to the stellar evolution calculations, zero-metal stars with $M_{\star}<0.9 M_{\odot}$ have a convective envelope \citep{Ri_2002}, where $M_{\star}$ is the mass of a star.
 The different internal structure from massive stars gives different thermodynamical properties of the stellar atmosphere. It is widely discussed that the surface convective motions excite  magnetohydrodynamic (MHD) waves and they transport the energy to the upper atmosphere to form the corona \citep{Ho_1982, KuSh1999, Ve_2010, Ma2016, SaSh2020, Va_2020}. The hot plasma in the corona emits the ultraviolet and soft X-ray radiations mainly from the closed magnetic loops. In fact, coronal activities have been detected by the X-ray radiation from  metal-poor stars as well as from the Sun \citep{FlTa1996,Ot_1997}.

The coronae of low-mass Pop III stars possibly give a significant contribution to the early cosmic reionization because they emit ionizing photons continuously for a sufficiently long duration due to the long lifetime. 
Recently, the dependence of coronal property on stellar metallicity has been investigated by MHD simulations for the coronal heating in open magnetic flux tubes \citep{Su2018} and closed magnetic loops \citep{WaSu2019}; both studies show that lower-metallicity stars give hotter and denser coronae because the radiative cooling is suppressed. It is also reported that zero-metal coronae emit strong UV and X-rays compared with the solar-metal coronae, which implies the importance of their radiative feedback at high-redshift.

However, it is still insufficient to understand the nature of metal-free coronae because no systematic survey has been conducted in a wide range of the parameters on magnetic flux tubes.
The numerical studies of the solar coronal loops have in fact reported that the geometric and physical parameters such as loop length, expansion factor and magnetic field strength have a crucial effect on the coronal properties \citep{An_2010, Da_2018}.
In particular, modeling the low-mass metal-free stars have large uncertainties in their magnetic properties.

The coronal activity is strongly connected with the stellar magnetic environment which is generated by the dynamo action in the surface convective layer. The dynamo activity is positively correlated with stellar rotation \citep{BrSu2005, Se_2017}. Although metallicity has a certain effect on the stellar dynamo \citep{Ka_2018}, there has been no robust relation established to describe the magnetic properties in zero/low-metal stars. Therefore we need to investigate the response of the coronal activities to various strengths and configurations of the magnetic field. We aim to understand systematic trends of low-mass Pop III stars with the magnetic properties and validate their radiative feedback in the early universe.

This paper has the following structure. In Section \ref{sec2}, we describe the models and settings of our simulations. In Section \ref{sec3}, we show the results about the dependence of the coronal properties on the magnetic field strengths and stellar masses. We discuss the results and the radiative feedback of low-mass Pop III stars in Section \ref{sec4} and summarise the paper in Section \ref{sec5}.

\section{Simulations}
\label{sec2}

\subsection{Settings}
\label{sec21}

We carried out one-dimensional magnetohydrodynamic (MHD) simulations of the heating of coronal loops in metal-free stars.
To find the systematic properties, we perform the simulations with stellar mass $M=0.8, 0.7, 0.6, 0.5 M_{\odot}$ and a wide range of the magnetic field strength.

The loop model and simulations are basically similar to those in \cite{WaSu2019};
we consider a single magnetic flux tube that is anchored at the photosphere and take the coordinate $s$ along the loop. We inject velocity perturbations $\delta v$ in three directions from the footpoints to excite MHD waves. 

It is assumed that the loop expands towards a higher altitude with an expansion factor, $f(s)$:
\begin{align}
%\begin{split}
f(s) = f_{\mathrm{max}} \times \frac{1}{2} [ \mathrm{tanh} \{ a (\frac{-|s|/R_{\star}+b}{h/R_{\star}}+\frac{\pi}{4})\} + 1],
%\end{split}
\label{eq1}
\end{align}
where $f_{\rm max}$ is the value of $f(s)$ at the loop top. $R_{\star}$ is the stellar radius and $h$ is the loop height. $a$ is a constant for each flux tube and we assume a functional form of $a=-0.54h/10^4 +0.6\times \mathrm{log}_2( f_{\rm max}/100) + 5.3$, where we set $b=0.02$.
The magnetic field along the loop is described as $B_s = B_{\rm ph}/f(s)$, where $B_{\rm ph}$ is the photospheric magnetic field strength.

We adopt the loop length $l=8\times 10^4$km, which gives $h \simeq 2.5\times 10^4$ km, for $M=0.8M_{\odot}$. We note that these are roughly comparable to moderate-sized solar coronal loops.  
The loop lengths for different stellar masses are set to be proportional to the scale height $\propto a_{\rm ph}^2/g$ where $a_{\rm ph}$ and $g$ are the sound speed and the gravitational acceleration at the photosphere, respectively (See Table \ref{table1}).

We solved the ideal MHD equations including thermal conduction and radiative cooling as follows;

\begin{align}
 \frac{\partial \rho}{\partial t} + \nabla \cdot \rho\bm{v} = 0,
 \label{eq2}
\end{align}
\begin{align}
 \rho\frac{\partial \bm{v}}{\partial t} = -\nabla \left(P + \frac{B^2}{8\pi}\right)+ \frac{1}{4\pi}\left (\bm{B} \cdot\nabla\right) \bm{B} - \rho\left(\bm{v} \cdot \nabla\right)\bm{v}- \frac{\rho GM}{R^2}\hat{R}, % \rm{sin}¥theta
\end{align}
\begin{align}
 \frac{\partial \bm{B}}{\partial t} = \nabla\times\left(\bm{v} \times \bm{B}\right)  ,
\end{align}

\begin{align}
\begin{split}
 &\rho\frac{d}{dt}\left(e+\frac{v^2}{2}+\frac{B^2}{8\pi\rho}- \frac{GM}{R^2} \right) +\nabla \cdot \left[\left(P+\frac{B^2}{8\pi}\right)v-\frac{B}{4\pi}\left(B\cdot\bm{v}\right)\right]  \\
 &+ \nabla\cdot F_c  +  q_R = 0
 \end{split}
 \label{eq5}
\end{align}
and

\begin{align}
P = \frac{\rho k_B T}{\mu m_u},
\label{eq6}
\end{align}
where $G$ is the gravitational constant and $R$ is the distance from the center of a star. $e$ is the internal energy, which has the relation $e=\frac{P}{(\gamma - 1)/\rho}$ where $\gamma = 5/3$. 
$F_c$ is the thermal conductive flux which has the form $F_c=\kappa T^{5/2} \frac{\partial T}{\partial s}$, where $\kappa$ is the Spitzer conductivity.
$\mu$ is the mean molecular weight that we adopt $\mu = 0.6$ for fully ionized plasma, $m_u$ is the atomic mass unit and $k_B$ is the Boltzmann constant.
Note that when we solve Equations (\ref{eq2})-(\ref{eq6}), it is taken into account the following curvature effects using Equation (\ref{eq1});
\begin{align}
\nabla \cdot \bm{X} = \frac{1}{f} \frac{\partial}{\partial s} f X_s,
\end{align}
\begin{align}
\nabla \times \bm{X} = \frac{1}{\sqrt{f}} \frac{\partial}{\partial s} \sqrt{f} X_s ,
\end{align}
where $\bm{X}$ and $X_s$ are any vector and its $s$ component. 

\begin{table*}
 \caption{Input parameters for our simulations. 
Each column presents stellar mass, stellar radius, effective temperature, photospheric density, standard magnetic field strength at the photosphere, magnitude of velocity perturbations, minimum and maximum periods for the $\delta v$ injection and loop length.}
  \centering
 \begin{tabular}{|c|c|c|c|c|c|c|c|c|c|c|} \hline
  $M [M_{\odot}]$ & $R_\star[10^5 \rm km]$& $T_{\mathrm{eff}}$[K] &$\rho_{\rm ph}[10^{-7} \rm g/cm^3]$ & $B_{\rm ph *} [\rm kG]$ & $\delta v [\rm km/s]$  & $\omega_{\rm max}^{-1}$ [s]& $\omega_{\rm min}^{-1}$[s] & $l[10^4 \rm km]$\\ \hline
  0.8 & 5.33 & 6356 & 4.88  & 2.32 & 1.14  &24 &2100& 8.00\\ \hline
  0.7 & 4.29 & 5836 & 10.39 & 3.24  & 0.787 &17 &1600& 5.42\\ \hline
  0.6 & 3.51 & 5336 & 23.20 & 4.63  & 0.535 &13 &1200& 3.87\\ \hline
  0.5 & 2.92 & 4793 & 57.46 & 6.91  & 0.343 &10 &900 & 2.88\\ \hline  \end{tabular}
  \label{table1}
   \end{table*}

$q_R$ is radiative cooling rate per volume. Since our simulations treat the wide range of atmospheric layers from the photosphere to the corona, we adopt different prescriptions for the chromosphere and the upper regions. 
In the chromosphere, we utilized the formulation by \cite{Su2018} which is extended from the empirical model for the solar atmosphere \citep{AnAt1989};
\begin{align}
q_R = 4.5 \times 10^9 \rho (0.2 + 0.8 \frac{Z}{Z_{\odot}}) \hspace{1.5mm} \rm{erg \hspace{1mm} cm^{-3} s^{-1}}
\label{eq9}
\end{align}
Equation (\ref{eq9}) is derived based on the observation of the solar chromosphere; 80 \% of the chromospheric emission is contributed from heavy elements. 

In the optically thin region, 

\begin{align}
q_R =
%\left\{
%\begin{array}{ll}
  \Lambda n n_e \hspace{1.5mm}  \rm{erg \hspace{1mm}cm^{-3} s^{-1}}. % \vspace{3mm} for optically thin \\ 
%  4.5 \times 10^9 \rho (0.2 + 0.8 \frac{Z}{Z_{\odot}})  \rm for optically thin
%\end{array}
%\right.
\label{eq10}
\end{align}
$\Lambda$ is the cooling function, for which we adopt from \cite{SuDo1993} for $Z=0$. $n$ and  $n_e$ are ion and electron number densities.

In addition, we define the cutoff temperature of radiative cooling, $T_{\rm off}$, such that $q_R =0$ when $T<T_{\rm off}$. It controls a minimum temperature of the low atmosphere, $T_{\rm min}$. Although $T_{\rm min}$ is often lower than $T_{\rm off}$ because the adiabatic cooling works even when $q_{\rm R}$ is turned off, $T_{\rm min}$ is generally comparable to $T_{\rm off}$. A classical model of the solar chromosphere gives $T_{\rm min}\approx 4000$ K $\approx 0.7 T_{\rm eff}$ \citep{Ve_1981}. Because $T_{\rm min}$ for low-metallicity stars has not been specified in observations, we simply refer to this classical solar value to set $T_{\rm off}=0.7T_{\rm eff}$.

We adopt the 2nd order Godunov method for compressible waves and the method of characteristics for incompressible waves \citep{StNo1992, SuIn2005}. 

\subsection{Input Parameters}
\subsubsection{Basic Stellar Parameters}

The stellar parameters for metal-free stars with different masses are shown in Table \ref{table1}.
Stellar radius $R_\star$ and effective temperature $T_{\rm eff}$ are adopted from stellar evolution calculations for $Z=0$ and stellar age $t=5$Gyr by \cite{Yi_2001,Yi_2003}.  The photospheric density, $\rho_{\rm ph}$, is derived by interpolating ATLAS model in \cite{Ku1979} using $M, R_\star$ and $T_{\mathrm{eff}}$.

The magnitude of velocity perturbation, $\delta v$, is determined by the surface convective flux; $\delta v \propto \rho_{\rm ph}^{-1/3} T_{\rm eff}^{4/3}$ \citep{St1988}.
We injected $\delta v$ with the power spectrum $P(\omega) \propto \omega^{-1}$, where $\omega$ is the frequency that has the range with $\omega_{\mathrm{min}}\leq \omega \leq  \omega_{\rm max}$. $\omega_{\mathrm{min, max}}$ has different values for the stars with different masses that are proportional to $g/a_{\mathrm{ph}}$.

\subsubsection{Magnetic Field}
\label{sec222}
\begin{figure*}
\begin{centering}
\includegraphics[width=180mm]{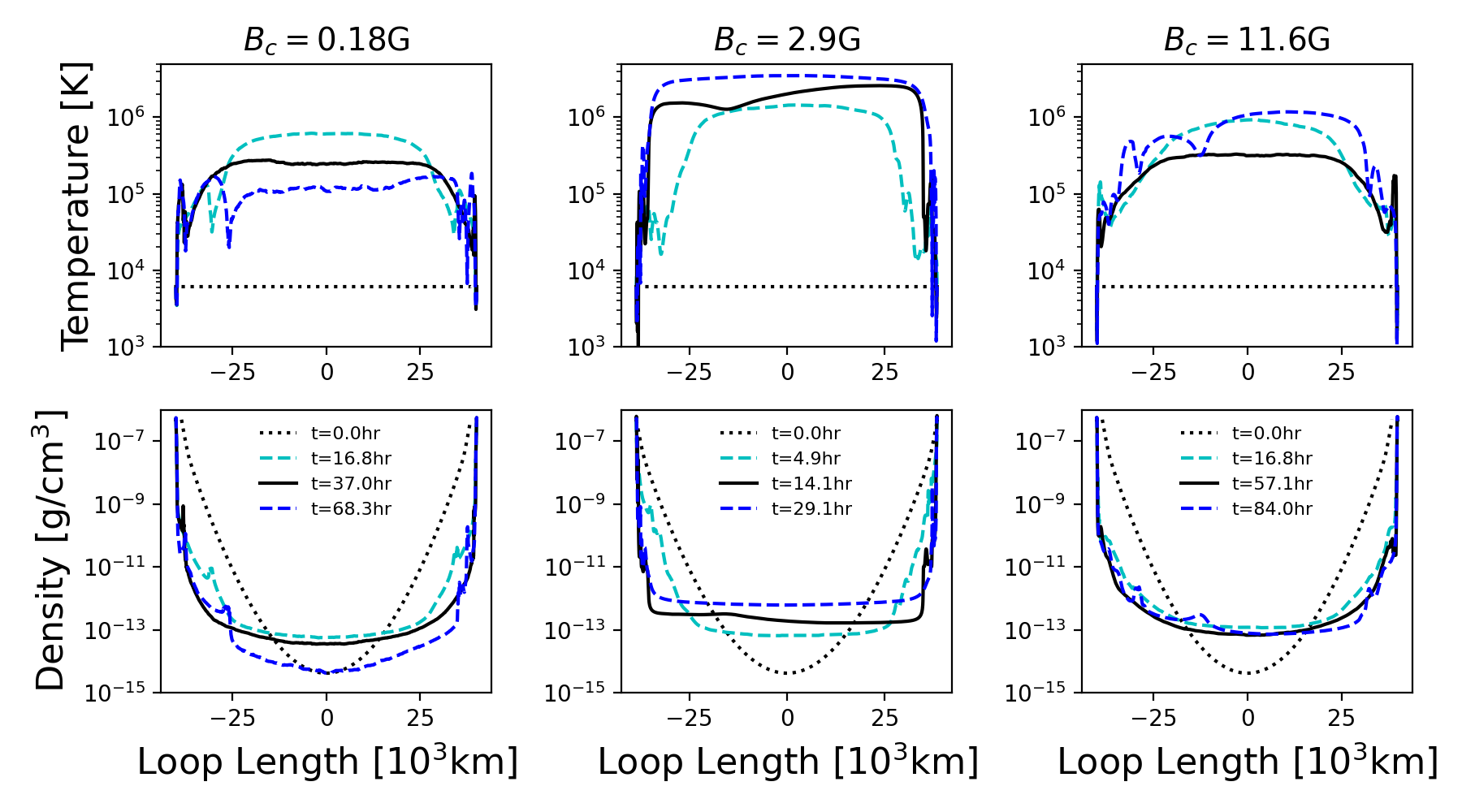}
\end{centering}
  \caption{Dynamical evolution of the loop structures for the cases of $M=0.8 M_{\odot}$ and $B_{\rm c}=0.18$ G (left), $2.90$ G (middle), and $11.6$ G (right). 
  The top and bottom panels show the temperature and density profiles.
  Line types correspond to different times that are shown in the bottom panels.
  }
  \label{fig1}
    \end{figure*} 

The magnetic activities in low-metallicity stars, not to mention metal-free stars, have large uncertainties because we have little information from observations. We therefore ran the simulations with various magnetic field strengths and investigate how the magnetic fields affect the coronal properties.

The photospheric magnetic field, $B_{\rm ph}$, is generally determined by magnetic amplification in the convective envelope, which is more effective for faster rotators.
It is difficult to specify $B_{\rm ph}$ for zero-metal stars, then we first set  $B_{\rm ph*}$ as the standard value which satisfies the condition of equipartition at the photosphere $\frac{8\pi B_{\rm ph*}^2}{P_0}=1$, where $P_0$ is gas pressure at the photosphere. $B_{\rm ph*}$ for stars with different masses is shown in Table \ref{table1}. To study the coronal properties under various magnetic environments, we carried out the simulations with the range of $B_{\rm ph}=[{\frac{1}{64},\frac{1}{32},\frac{1}{16},\frac{1}{8},\frac{1}{4},\frac{1}{2},1, 2, 4, 8}]\times B_{\rm ph*}$. 

The expansion factor of magnetic flux tubes is also an important parameter because it determines the magnetic field strength in the upper atmosphere.
We vary $f_{\rm max}$ in the range of $100 \leq f_{\rm max} \leq 3200$, where we increase the value by a factor of two up to $f_{\rm max}=400$, and then every 400 thereafter.

Once $B_{\rm ph}$ and $f_{\rm max}$ are fixed, magnetic field in the corona, $B_{\rm c}=B_{\rm ph}/f_{\rm max}$, is determined, which corresponds to the average strength of the open magnetic field regions. In this study, we mainly focus on the dependence of coronal properties on $B_{\rm c}$.
The value of $B_{\rm c}$ is fixed by the combination of $B_{\rm ph}$ and $f_{\rm max}$. The range of $B_{\rm c}$ for each stellar mass is $4.5\times 10^{-2}$G $\leq B_{\rm c}\leq 23.2$ G for $M=0.8M_{\odot}$, $6.3\times 10^{-2}$G $\leq B_{\rm c}\leq 64.8$ G for $M=0.7 M_{\odot}$, $6.0\times 10^{-2}$G $\leq B_{\rm c}\leq 55.8$ G for $M=0.6 M_{\odot}$ and $9.0\times 10^{-2}$G $\leq B_{\rm c}\leq 69.1$ G for $M=0.5 M_{\odot}$.

The grid length is set to resolve the shortest wavelength of the input waves by at least 10 grid points; the minimum grid length, $\Delta s_{\rm min}$, and the maximum grid length, $\Delta s_{\rm max}$, depend on $B_{\rm ph}$, and the scale height and $f_{\rm max}$, respectively. They follow the relations below;
\begin{align}
\Delta s_{\rm min} = \mathscr{A} R_* \left( \frac{B_{\rm ph}}{B_{\rm ph*}}\right),
\end{align}
\begin{align}
\Delta s_{\rm max} = \mathscr{B} R_* \left(\frac{g}{a_{\rm ph}^2}\right)f_{\rm max}^{-1},
\end{align}
where $\mathscr{A}=2.23\times 10^{-5}$ and $\mathscr{B}=8.0$ are constants.
For instance, $\Delta s_{\rm min} \approx 12$km and $\Delta s_{\rm max}\approx 180$km for the case with $M=0.8M_{\odot}$, $B_{\rm ph}=B_{\rm ph*}$ and $f_{\rm max}=200$.
   
\section{Results}
\label{sec3}
\subsection{Time Evolution of the Loop Profiles}
\label{sec31}

In our model, the initial cool gas with $T=T_{\rm eff}$ is heated via the  dissipation of Alfv\'{e}n waves. In our simulations, the main channel of the wave dissipation is the nonlinear mode conversion from incompressible Alfv\'{e}nic waves to compressible waves \citep{SuIn2005,SuIn2006}). In addition, fast-mode MHD shocks, which is formed by the direct steepening of transverse (Alfv\'{e}nic) waves \citep{Hol1982, Su2004}, is another channel of the wave dissipation.

 Figure \ref{fig1} shows the time evolution of the loop profiles for the cases of $M=0.8M_{\odot}$ and $B_{\rm c}=0.18, 2.9$ and $11.6$ G.
 As can be seen from the figure, most of our simulations indicate that the simulated coronal loops undergo the dynamical evolution with intermittent rises and drops of the temperature (top panels). Such thermally non-equilibrium behaviors have been actually seen in observations \citep{Au_2014,Fr_2015} as well as simulations \citep{Mu_2003, Mu_2004,An_2010,Fr_2018}. However, the dynamical behavior in our simulations may be partially because of our limited treatment of the wave heating in 1D flux tubes (see Section \ref{sec42}).
%because the phenomenon depends on the heating distribution and the loop geometry (Muller et al. 2003; Antolin and Shibata 2010; Froment et al. 2018).

The left panels in Figure \ref{fig1} correspond to the case with weak $B_{\rm c}=0.18$ G.
In this case, the corona in excess of $T=10^6$K is not formed. It is only heated up to $T\sim (3-5)\times 10^5$K and fluctuates below $10^6$K .
Smaller $B_{\rm c}$ leads to the larger nonlinearity of Alfv\'{e}nic waves $\propto \delta B /B_s$, which enhances the wave dissipation. Therefore, a larger fraction of the Alfv\'{e}nic waves dissipate at the lower altitude where the density is higher and most input energy is lost through radiative cooling.
%The heating at denser regions is more efficiently lost through radiative cooling, rather than increasing the temperature.
In contrast, only a smaller fraction of the initial wave energy remains at higher altitudes, and therefore, the upper atmosphere is not heated up to the coronal temperature.

It is also useful to evaluate characteristic timescales.
We define the heating timescale $\tau_{\rm heat}$ as the travel time required for the Alfv\'{e}n wave to damp the Poynting flux $F_{\rm A}$ by a factor of e:%attenuate the input Poynting flux $F_{\rm A,0}\approx \rho_{\rm ph} \delta v^2 v_{\rm A}$ by a factor of 1/e;
\begin{align}
\tau_{\rm heat} = \frac{1}{v_{\rm A}}\frac{ds}{d\ln F_{\rm A}},
%\label{eq13}
\end{align}
where we evaluate $v_{\rm A}$ and $F_{\rm A}$ at the loop top, which yields $\tau_{\rm heat}=250$ s for this case.
The radiative cooling timescale is given by 
\begin{align}
\tau_{\rm cool} = \frac{\rho e}{n^2 \Lambda} .
\label{eq14}
\end{align}
We note that the dependence of $\tau_{\rm cool}$ on $T$ is different for different temperature ranges; for $T>10^6$ K, $\Lambda \propto T^{1/2}$ so that $\tau_{\rm cool} \propto T^{1/2} \rho^{-1}$. On the other hand, for $T<10^6$ K, $\Lambda \propto T^{-1}$ in the case of zero metallicity \citep{SuDo1993, WaSu2019} and thus $\tau_{\rm cool} \propto T^2 \rho^{-1}$.
For both temperature ranges, $\tau_{\rm cool}$ increases with temperature so that the gas should be thermally unstable \citep{Ba1995}. However, for $T>10^6$ K, the gas is not so unstable because the dependence on $T$ is relatively weak and the thermal conduction $\propto T^{5/2}$ works to stabilize it. In contrast, for $T<10^6$ K, the gas is quite unstable owing to the inefficient thermal conduction and strong dependence of $\tau_{\rm cool}$ on $T$.

If we estimate $\tau_{\rm cool}$ for the time-averaged values of $T=2.5\times 10^5$K and $\rho =4\times 10^{-14}$ g cm$^{-3}$ at the loop top, we obtain $\tau_{\rm cool}=180$ s. 
The comparison between $\tau_{\rm heat}$ and $\tau_{\rm cool}$ indicates that the heating is dominated by the radiative cooling in the weak $B_{\rm c}$ case. Therefore, high-temperature corona is not formed in this case.

In the case of intermediate $B_{\rm c}=2.9$ G (the middle panels), the corona is rapidly heated up to $T = 10^6$K and reaches $T\sim 3\times 10^6$K at $t=29.1$hr.
In this case, the wave nonlinearity, $\delta B / B_s$, is moderately smaller  and a larger fraction of the input Alfv\'{e}nic waves reach the corona to heat the gas.
The heating timescale is $\tau_{\rm heat}=310$ s, which is consistent with the slower dissipation compared with the weak $B_{\rm c}$ case. The cooling timescale is $\tau_{\rm cool}=480$ s, where we use the time-averaged loop top values of $T=2.2 \times 10^6$K and $\rho=2.0 \times 10^{-13}$ g\hspace{1mm}cm$^{-3}$. $\tau_{\rm cool}\propto T^{1/2}$ is long enough owing to the high temperature, and consequently the cooling is dominated by the heating, $\tau_{\rm cool} > \tau_{\rm heat}$.   Therefore,
%The hotter corona induces more efficient chromospheric evaporation and the considerably higher coronal density is obtained. 
the loop profile is kept almost steady with time once the temperature is in the thermally stable region of $T>10^6$K.

The right panels show the case with strong $B_{\rm c}=11.6$ G.
The corona is heated up to $T\sim 10^6$K, however, it does not settle into the stable state. 
In this case, $\tau_{\rm heat}=440$ s and $\tau_{\rm cool}=310$ s using $T=7\times 10^5$K and $\rho=1.0 \times 10^{-13}$ g\hspace{1mm}cm$^{-3}$ for the time-averaged values. However, it is found that $\tau_{\rm cool}$ fluctuates widely over time; the gas is efficiently heated at early time and it leads more chromospheric evaporation to increase the coronal density. 
 Then, the cooling time drops to $\tau_{\rm cool}\sim 1$ s and the denser gas suffers more effective radiative cooling so that the temperature drops to the thermally unstable range of $T\lesssim 5\times 10^5$K. As the loop structure collapses and the density also decreases, $\tau_{\rm cool}$ increases to $\sim 10^3$ s and the corona is reheated to $T \sim 10^6$K. The loop structure evolves dynamically with time in the temperature range $10^5$K $<T<10^6$K in a quasi-periodic manner.

\begin{figure*}
\begin{centering}
\includegraphics[width=180mm]{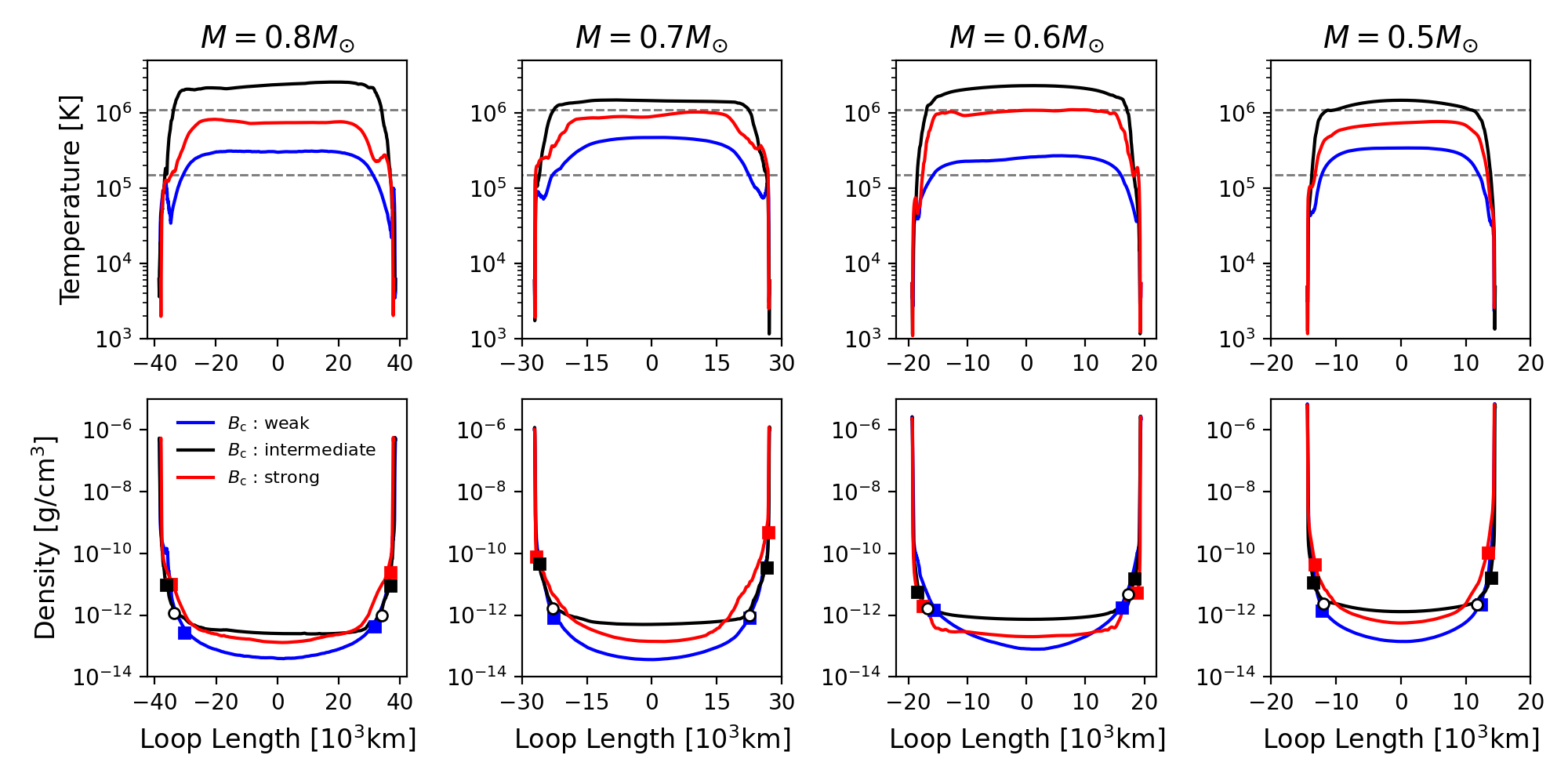}
\end{centering}
  \caption{Comparison of the time-averaged temperature and density structures for different $M$ and $B_{\rm c}$. The line color indicates the different $B_{\rm c}$; weak (blue), intermediate (black) and strong (red) cases. The gray dashed lines in the upper panels denote  $T=1.5\times 10^5$K and $T=1.1 \times 10^6$K. The filled squares and open circles in the bottom panels correspond to the locations where $T=1.5 \times 10^5$K and $1.1\times 10^6$K, respectively. }
  \label{fig2}
\end{figure*}  

The more dynamic nature of stronger $B_{\rm c}$ cases is also interpreted in terms of the intermittency of the wave heating.
The heating in the loops are done by shock waves that are nonlinearly excited from Alfv\'{e}nic waves. The wavelength of the Alfv\'{e}nic waves is longer for larger $B_{\rm c}$ owing to the faster Alfv\'{e}n velocity, which also gives a longer spatial interval between neighboring shocks.
The shock heating does not occur in a continuous manner but at localized regions in a stochastic manner in stronger $B_{\rm c}$ cases.  
Therefore the stable hot corona is hardly kept but the temperature goes up and down in a cyclic manner. 

\subsection{Dependence of Coronal Properties on \texorpdfstring{$B_c$ and $M$}{Bc and M}}
\label{sec32}

To compare the average properties of loops with different $B_{\rm c}$ and $M$, we show the time-averaged distributions of temperature and density in Figure \ref{fig2}.
The profiles for different $B_{\rm c}$ are displayed by different line colors in each panel; the blue, black and red lines correspond to weak ($<1$ G), intermediate ($\geq 1$ G and $\leq 10$ G) and large ($>10$ G) $B_{\rm c}$, respectively.  The parameter sets $(B_{\rm ph}/B_{\rm ph*}, B_{\rm c})$ for the different cases are shown in Table \ref{table2}.

\begin{table}
\caption{Parameter sets for the cases in Figure \ref{fig2}. }
 \centering
 \begin{tabular}{c|| c c c} \hline
  $M [M_{\odot}]$ & &($B_{\rm ph}/B_{\rm ph*}, B_{\rm c}$)& \\\hline
    & weak & intermediate & strong \\ \hline
  0.8 & (1/64, 0.18) & (1, 2.90) & (2, 11.6) \\
  0.7 & (1/64, 0.25) & (1, 4.05) & (4, 32.4) \\ 
  0.6 & (1/64, 0.36) & (2, 4.63) & (4, 23.2) \\ 
  0.5 & (1/64, 0.54) & (2, 5.76) & (4, 23.0) \\ \hline 
  \end{tabular}
  \label{table2}
   \end{table}

\begin{figure*}
\begin{centering}
\includegraphics[width=180mm]{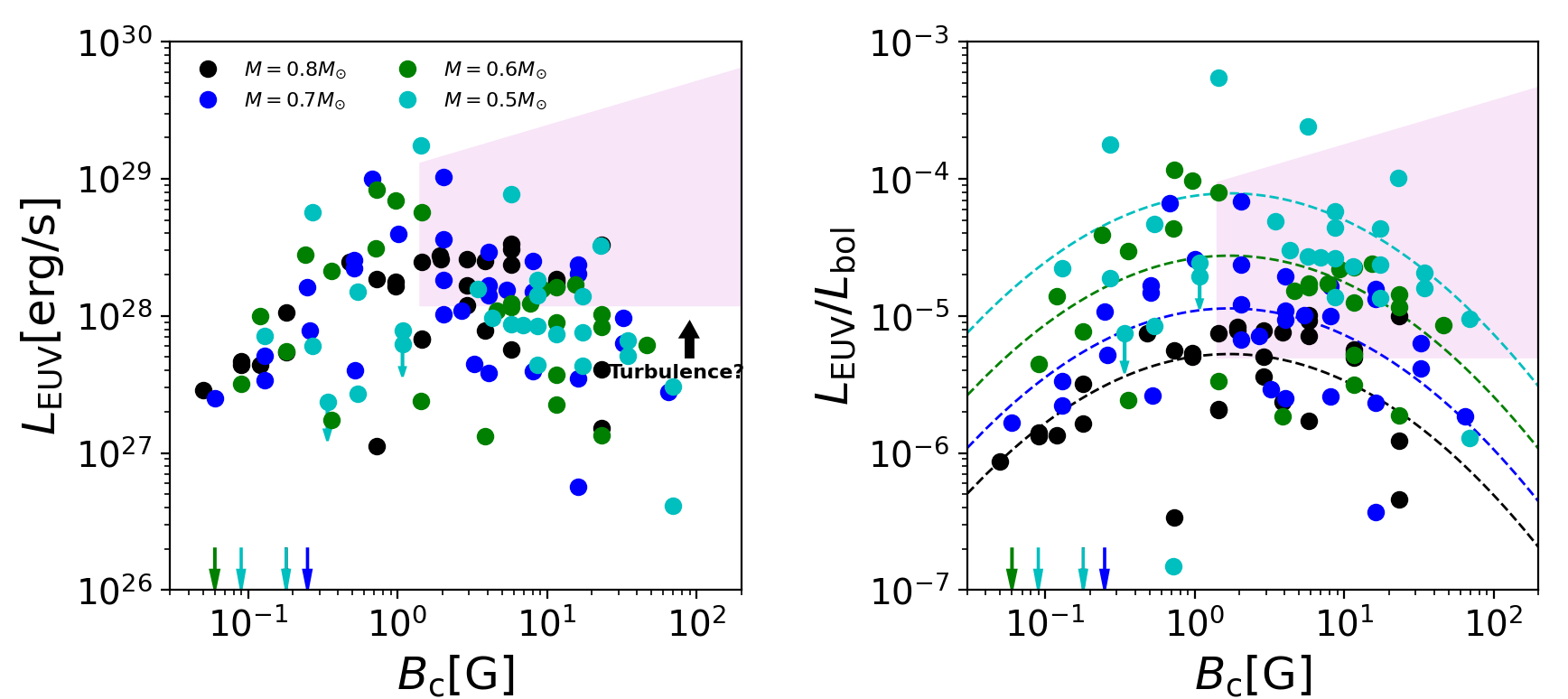}
\end{centering}
  \caption{Left panel: EUV luminosity $L_{\rm EUV}$ with $B_{\rm c}$. Colors of the points show the different stellar masses. The points with arrow are the upper limits. The cases that give low $L_{\rm EUV}$ outside the vertical range are indicated by single arrows at the bottom side of the panel. The pink shaded region is a speculated tendency if the turbulent heating process is active (see text). Right panel: $L_{\rm EUV}/L_{\rm bol}$ with $B_{\rm c}$. The points are plotted in the same manner as in the left panel. The dashed lines are the fitted curves of Equation (\ref{eq16}). }
\label{fig3}    
\end{figure*}  
\begin{figure*}
\begin{centering}
\includegraphics[width=180mm]{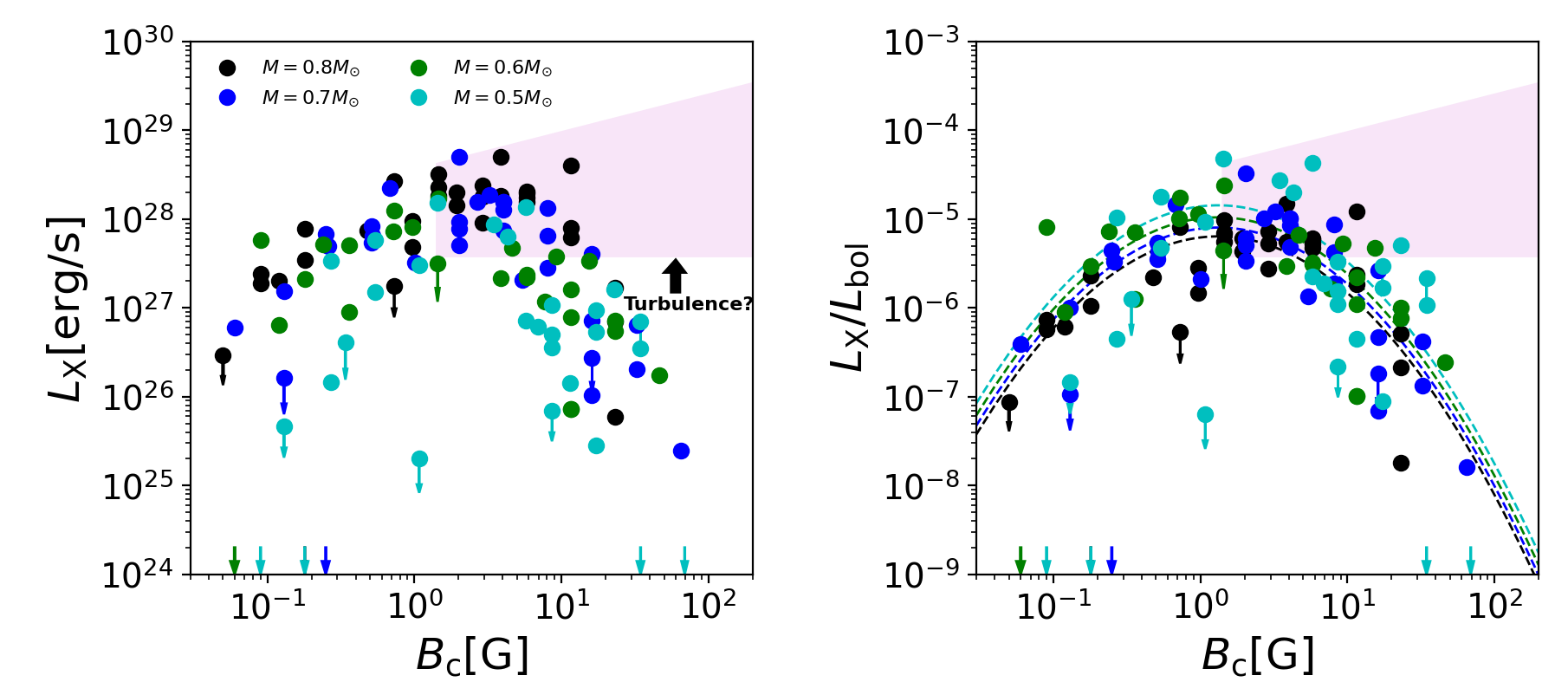}
\end{centering}
  \caption{Same as Figure \ref{fig3}, but for $L_{\rm X}$.
  The dashed lines in the right panel are the fitted curves of Equation (\ref{eq17})}
  \label{fig4}
\end{figure*} 
\begin{figure*}
\begin{centering}
\includegraphics[width=180mm]{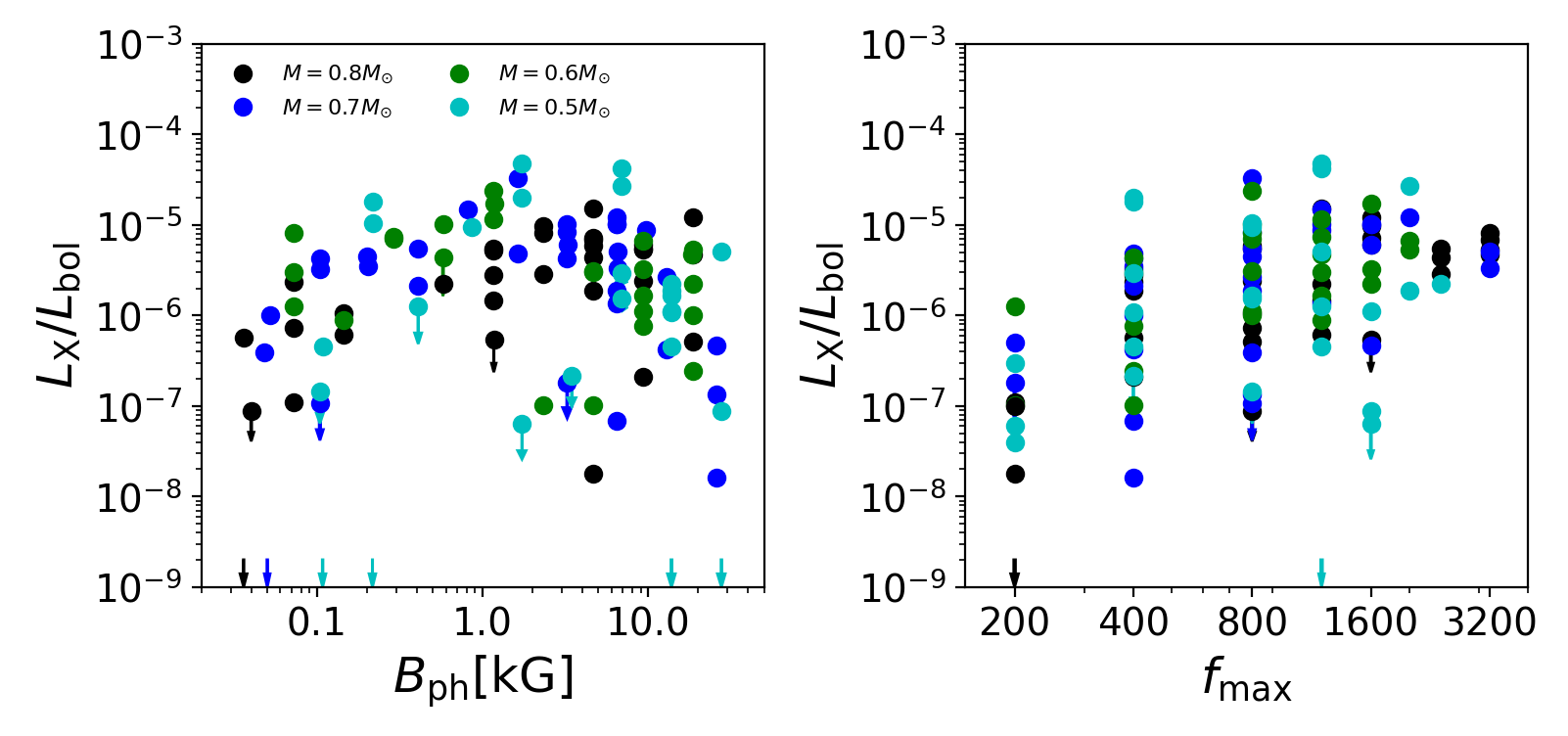}
\end{centering}
  \caption{$L_{\rm X}/L_{\rm bol}$ with $B_{\rm ph}$ (left panel) and $f_{\rm max}$ (right panel).}
  \label{fig5}
\end{figure*}  

The dependence on $B_{\rm c}$ show a similar tendency for different masses: the hottest and densest corona is obtained in the cases with the intermediate $B_{\rm c}$.
For the smaller $B_{\rm c}$, the fast dissipation in the lower atmosphere makes it difficult to form the hot corona. For the larger $B_{\rm c}$, the higher density enhances the radiative cooling and the coronal temperature of $T>10^6$K is hard to be maintained for a long time. Thus, the loop dynamically evolves with time and the time-averaged temperature and density are lower than those of the intermediate cases. However, we note that the behavior of the cases with strong coronal field, $B_{\rm c}>10$ G may be modified if we consider additional heating processes, such as the turbulent cascade of Alfv\'{e}nic waves, which is not taken into account in our simulations. We will discuss this issue in Section \ref{sec42}.

When we focus on the same line colors in each panel, it can be seen that the density is higher for lower-mass stars. The gas pressure at the photosphere is larger for lower-mass stars. In addition, the loop height, which is proportional to the pressure scale height (Section \ref{sec2}), is shorter for lower-mass stars. Therefore, the coronal density of lower-mass stars is higher even though the difference between the photospheric density and the coronal density is higher.

\subsection{EUV and X-ray Luminosities}
\label{sec33}

We estimated the radiative fluxes in extreme ultraviolet (EUV) and X-ray wavelengths to understand the correlation between the coronal magnetic field and the high-energy radiation. We here assume that the star is covered by the same simulated loops and calculate the luminosity $L$ as follows;
\begin{align}
&L = \frac{1}{f_{\rm max} } \times 4 \pi R_{\star}^2 \int q_R f(s) ds .
\end{align}
%\begin{align}
%&L_X =  \frac{1}{f_{\rm top} } \times 4 \pi R^2 \int_{s(T > 1.1 \times 10^6 K)} q_R f(s) ds  .
%\end{align}
We label $L_{\rm EUV}$ and $L_{\rm X}$ for $L$ in the temperature ranges of  $1.5\times 10^5 \mathrm{K} \leq T \leq 1.1\times 10^6$K and $T > 1.1\times 10^6$K, respectively. We note that $T=1.5\times 10^5$K and $=1.1\times 10^6$K respectively correspond to 13.6 eV and 0.1 keV.  

The left panel of Figure \ref{fig3} shows the time-averaged  EUV luminosities $L_{\rm EUV}$ of all the simulated data against $B_{\rm c}$.
We note that multiple cases with different sets of the input parameters of $B_{\rm ph}$ and $f_{\rm max}$ (Section \ref{sec222}) give the same $B_{\rm c}$ for each $M$.
$L_{\rm EUV}$ is peaked around $B_{\rm}\approx$ a few G, which is consistent with Figure \ref{fig2} and the discussion in Section \ref{sec32}; the cases with intermediate $B_{\rm c}=2-6$ G give the hot and dense coronae. Toward weaker $B_{\rm c} < 1$ G and stronger $B_{\rm c} > 10$ G, $L_{\rm EUV}$ gradually decreases. In the top panels of Figure \ref{fig2}, the horizontal dashed lines indicate $T=1.5\times 10^5$K and $T=1.1\times 10^6$K, and we define that the EUV radiations are from the plasma between these temperatures. The filled boxes and the open circles in the bottom panels point the location of these temperatures. Figure \ref{fig2} illustrates that, although the hot corona is not formed in small and large $B_{\rm c}$ cases, EUV radiations are emitted from broad regions of the loops in these cases. 

Some cases occasionally emit quite strong radiation although the average temperature around the loop top is low $\lesssim 5\times 10^5$ K.  We show the upper limits of these cases in the circles with arrows in Figure \ref{fig3}.
 Some simulated cases with weak $B_{\rm c}$ give quite small $L_{\rm EUV}$, which is below the displayed range of $L_{\rm EUV}\ge 10^{26}$erg s$^{-1}$. These cases, which are indicated by the single arrows in the bottom left side of the panel, imply that there is a condition for the coronal magnetic field strength to emit sufficiently large $L_{\rm EUV}$. 

The right panel of Figure \ref{fig3} shows the EUV luminosity normalised by the bolometric luminosity, $L_{\rm EUV}/L_{\rm bol}$, with $B_{\rm c}$. 
We derived $L_{\rm bol}$ from $L_{\rm bol}=4\pi R_\star ^2 \sigma T_{\rm eff}^4$, where $\sigma$ is the Stefan-Boltzmann constant and we use the values of $R_\star$ and $T_{\rm eff}$ from Table \ref{table1}.  In this panel, the dependence on stellar mass is clearly visible; lower-mass stars give higher $L_{\rm EUV}/L_{\rm bol}$ because the gas density in the loops are higher. The peak value, $L_{\rm EUV}/L_{\rm bol} > 10^{-4}$ for  $M=0.5M_{\odot}$, while $L_{\rm EUV}/L_{\rm bol} \sim 8\times 10^{-6}$ for $M=0.8M_{\odot}$.

%For $B_{\rm c}<1$G, $L_{\rm EUV}$ increases with $B_{\rm c}$.
%On the other hand, for $B_{\rm c}>1$G, $L_{\rm EUV}$ decreases with $B_{\rm c}$. In this region, the variation of $v_{\rm A}$ gets large between photosphere and corona, and less waves reach the corona with increasing $B_{\rm c}$.

Figure \ref{fig4} shows the time-averaged X-ray luminosities $L_{\rm X}$ and $L_{\rm X}/L_{\rm bol}$ with $B_{\rm c}$.
The trend is almost similar to that of $L_{\rm EUV}$, while it has a strong dependence on $B_{\rm c}$ compared with $L_{\rm EUV}$. 
$L_{\rm X}$ decreases toward both weaker and stronger sides of $B_{\rm c}$. In the weak $B_{\rm c}$, the X-ray corona with $T\ge 1.1\times 10^6$K is hardly heated up (Figures \ref{fig1} \& \ref{fig2}). 
In the strong $B_{\rm c}$, on the other hand, although the X-ray corona is formed transiently, the temperature drops in a quasi-periodic manner. Therefore, the time-averaged $L_{\rm X}$ is lower than those of intermediate $B_{\rm c}$ cases. 

$L_{\rm X}/L_{\rm bol}$ is higher for lower mass as with $L_{\rm EUV}/L_{\rm bol}$.
At the peak, $L_{\rm X}/L_{\rm bol} \sim 5\times 10^{-5}$ for  $M=0.5M_{\odot}$, while $L_{\rm X}/L_{\rm bol} \sim 10^{-5}$ for $M=0.8M_{\odot}$.
Some cases with $M\ge 0.7 M_{\odot}$ emit strong X-rays, $L_{\rm X}/L_{\rm bol} \gtrsim L_{\rm EUV}/L{\rm bol}$. 

We additionally show the dependence of $L_{\rm X}/L_{\rm bol}$ on the other magnetic parameters of $B_{\rm ph}$ and $f_{\rm max}$, in Figure \ref{fig5}. 
It can be seen that both are related to $L_{\rm X}/L_{\rm bol}$; the value of $L_{\rm X}/L_{\rm bol}$ tends to decrease at smaller and larger sides of $B_{\rm ph}$ and $f_{\rm max}$. However, the scatters on the same $B_{\rm ph}$ or $f_{\rm max}$ are larger than those on the same $B_{\rm c}$ (Figure \ref{fig4}).
Therefore, the dependence on $B_{\rm ph}$ and $f_{\rm max}$ is relatively weak, which indicates that $B_{\rm c}$ is a critical parameter that determines the radiative property of metal-free coronae.

Finally, we fitted the dependence of $L_{\rm EUV}/L_{\rm bol}$ and $L_{\rm X}/L_{\rm bol}$ on $B_{\rm c}$;
\begin{align}
\begin{split}
\mathrm{log}\left(\frac{L_{\mathrm{EUV}}}{L_{\mathrm{bol}}}\right) =& -5.85-5.75\mathrm{log}\left(\frac{M}{M_{\odot}}\right) \\
&-0.33(\mathrm{log}B_{\rm c})^2+0.16 \mathrm{log}B_{\rm c},
\end{split}
\label{eq16}
\end{align}
\begin{align}
\begin{split}
\mathrm{log}\left(\frac{L_{\mathrm{X}}}{L_{\mathrm{bol}}}\right) =& -5.37 -1.72\mathrm{log}\left(\frac{M}{M_{\odot}}\right) \\
&-0.82(\mathrm{log}B_{\rm c})^2 + 0.20 \mathrm{log}B_{\rm c},
\end{split}
\label{eq17}
\end{align}
where $M$ is the stellar mass.
The dashed lines in the right panels of Figures \ref{fig3} and \ref{fig4} indicate the fitted curves for the different masses.
$L$ is scattered on the same $B_{\rm c}$ because of the difference of the field strength, $B_{\rm ph}$, at the photosphere, however, the fitting formulae using only $B_{\rm c}$ roughly explain the tendency.
Again, we mention that the coronal properties at $B_{\rm c}>10$ G may be modified when other heating processes are taken into account. Therefore, the trends of $L$ in the strong $B_{\rm c}$ are not definite, which will be discussed in the next section.
    
\section{Discussion}
\label{sec4}
\subsection{Active Young Stars}
\label{sec41}

Low-mass main sequence stars with surface convection show a large variety of magnetic activity involving coronae and flares. Young solar-type stars generally exhibit brighter and hotter X-ray emissions \citep{Gu_1997, Ri_2005}. For example, EK Dra (HD129333) is a young solar analogue with age $\lesssim 0.1$ Gyr and gives the X-ray luminocity, $L_{\rm X}\simeq10^{30}$erg s$^{-1}$, or  log$\left(L_{\rm X}/L_{\rm bol}\right)\simeq-3.6$, which are about $10^3$ times more luminous than those of the Sun \citep{Gu_1995,Te2005}.

The X-ray luminosity is correlated with stellar rotation period \citep{Pa_1981, Pi_2003, Wr_2011}, which generally decreases with the stellar age (e.g., \citealp{Sk1972}). The solar-type stars follow the relation, $L_{\rm X} \approx (3\pm 1)\times 10^{28} t^{-1.5\pm 0.3}$erg/s, where $t$ is the stellar age in Gyr with a saturated luminosity $L_{\rm X}\leq 4\times 10^{30}$ erg/s \citep{Gu2007}. 
The strong X-rays from these (near) saturated stars probably originate from explosive flares that involve large loops \citep{ShYo1999}, in addition to the steady quiet corona as modeled in the present paper. 
Indeed, a large flaring loop whose height is approximately the stellar diameter has been directly observed in the Algol system \citep{Pe_2010}.

In our calculations, we fix the loop length for each stellar mass (Table \ref{table1}) by extrapolating that of typical solar coronal loops. Also, our model cannot treat magnetic reconnection that is believed to trigger flares \citep{Ts1996, Sh1999, Ta_2012}. In more realistic treatments, $L_{\rm X}$ from low-mass Pop III stars is expected to show a larger variety.   In addition, observational analysis with the large number of samples reports that the stellar spin evolution depends on metallicity and lower-metallicity stars rotate faster on average than higher-metallicity stars \citep{AmMa2020}. 
It supports that young low-mass Pop III stars have a potential to emit much brighter X-rays for a long duration than the simulated values shown in the previous section.

\subsection{Process of Wave Dissipation}
\label{sec42}

In our simulations Alfv\'{e}nic waves dissipate only by the excitation of compressible-mode waves, which is one of the limitations of our one dimensional treatment.  
In reality, however, other processes of the wave dissipation are also expected to play a role in the heating of the plasma, such as Alfv\'{e}nic turbulence \citep{Ma_1999, Cr_2007, Sh_2019}, phase mixing \citep{HePr1983, OfAs2002, MaSu2012}, and resonant absorption \citep{Io1978,An_2015}. 
In order to incorporate these processes we have to handle 2D/3D simulations or adopt additional models, which we plan to study in our future work. 

We here discuss Alfv\'{e}n wave turbulence, which is one of the reliable mechanisms that heat up the corona. Upgoing Alfv\'{e}nic waves interact with reflected backward propagating waves, which triggers cascading Alfv\'{e}n wave turbulence. This is an efficient mechanism of the dissipation of Alfv\'{e}nic waves. 
Recent studies report that both the shock and turbulent heating processes are essential in driving the solar wind \citep{Cr_2007, MaSu2014, Sh_2018, Ma2020} and coronal loops \citep{Ma2016}; the shock heating is dominant in the lower atmosphere, while the turbulent heating is dominant above the transition region. 

In our simulation with $B_{\rm c}>10$ G, a larger fraction of the input energy of the Alfv\'{e}nic waves travels into the upper atmosphere owing to the smaller nonlinearity $\delta B/B_s$.
% which the large background field, $B_s$, favors it \citep{PeCh2013, vaAs_2017, Sh_2018}. 
Therefore, the hot and dense corona is expected to be maintained in the cases with larger $B_{\rm c} > 10$ G by the contribution from the turbulent cascade of the Alfv\'{e}nic waves; we might underestimate $L_{\rm EUV}$ and $L_{\rm X}$ at $B_{\rm c}>10$ G. In contrast, the trend in the regime of $B_{\rm c} < 1$ G would not be affected, even if the turbulent heating was active, because most of the input Alfv\'{e}nic waves are damped before reaching the upper atmosphere.  
To summarise, we speculate that $L$ does not decrease at $B_{\rm c}<10$G and it may be flat or possibly increases as presented by the pink shaded regions in Figure \ref{fig3} and \ref{fig4}. It is required to incorporate the turbulent heating process to verify the tendency for our future work.

\subsection{Treatment of Chromosphere}
\label{sec43}

The chromosphere is an interface that transfers the energy from the photosphere to the corona. The temperature in the chromosphere determines the distribution of the density, which controls the propagation and dissipation of waves. The thermal properties of the chromosphere affects the heating of the upper layers. We adopt the simplified treatment for the radiative cooling in the chromosphere of metal-free stars (Equation (\ref{eq9})). The cooling rate is a simple extension from the solar value \citep{AnAt1989}, which is modeled by the empirical temperature profile of the VAL model C under the assumption of the static chromosphere with spatially uniform radiative emissions \citep{Ve_1981}.
However, the solar chromosphere has been known to be quite dynamic \citep{Ka_2007, De_2014, Sk_2015} and the local thermodynamic equilibrium (LTE) is not satisfied in general. In such circumstances,  it is required to solve radiative transfer and detailed rate equations in time-dependently to determine the radiative cooling rate.  

The dynamic chromosphere has been studied by theoretical models and numerical simulations \citep{CaSt2002,RaUl2003} and it is claimed that dynamic models show quite different radiative properties from a static model \citep{CaSt2002}. \cite{CaLe2012} introduced an approximated recipe to estimate the chromospheric emission from their radiative transfer calculation. A future direction for our study is to extend their recipe to chromosphers with different metallicities.  

The cutoff temperature of the radiative cooling, $T_{\rm off}=0.7\times T_{\rm eff}$, is also an important parameter because it controls the minimum temperature of the atmosphere. Smaller $T_{\rm min}$ leads to the more rapid decrease of the density, which enhances the reflection of Alfv\'{e}nic waves.  The modified treatment may give better understandings of $T_{\rm min}$. 

We additionally mention the effect of the magnetic diffusion.
The chromosphre is partially ionized and the collision between neutrals and charged particles causes the ambipolar diffusion.  It facilitates damping the Alfv\'{e}nic waves and heating the gas particularly in the upper chromosphere \citep{Si_2011, KhCo2012, Ma_2017}. The ambipolar diffusivity is described as $\eta_{\rm ambi} = (|B|\rho_n/\rho)^2/\rho_i \nu_{in}$, where $\rho_n, \rho_i, \rho$ and $\nu_{in}$ are neutral density, ion density, total density and ion-neutral collision frequency \citep{Ma_2015}. Although recent numerical simulations of the solar atmosphere indicate that ambipolar diffusion has a small effect on the chromospheric heating compared with the shock heating which is resulted from the nonlinear mode coupling \citep{Ar2016, BrAr2016}, it may be important under the metal-deficient environment; the metal-free condition lowers $\rho_i$ because of the  lack of the ionization sources and raises $\eta_{\rm ambi}$. Therefore, the effect of amibipolar diffusion would be large and needs to be added in the induction equation for proper modeling of metal-free chromospheres.

\subsection{Radiative Feedback by Low-mass Pop III Coronae}
\label{sec44}

The role of low-mass Pop III stars has not been investigated in detail in the context of the cosmic reionization to date, partly because the existence of low-mass Pop III stars is unclear and the high-energy radiation from them is quite uncertain.  In this subsection, we discuss the effect of the EUV and X-ray radiations from low-mass Pop III stars on the cosmic reionization in comparison to other ionizing sources.

One of the unique characteristics of low-mass Pop III stars is that they also emit soft X-rays, in addition to the EUV radiation. Our simulations show that their X-ray luminosity is roughly comparable to the EUV luminosity (Section \ref{sec33}), which is in distinctive contrast to  massive metal-free stars. 
The X-ray photons also play a role in partially ionizing the neutral gas. Previous studies report that the X-ray sources can contribute a few per cent of the ionization \citep{Oh2001, Mi_2011}. Furthermore they can heat up intergalactic medium (IGM hereafter) in a more spatially uniform manner due to their long mean path \citep{Ma_2004, Ri_2004, Mc2012}. X-ray binaries (XRBs) and mini quasars have been considered to be likely sources of the X-ray reionization.  In addition to these candidates, we consider whether the X-rays from low-mass Pop III stars can contribute to the process of the reionization.

We estimate the X-ray luminosity per unit star-formation rate (SFR) to directly compare to previous works that discuss other X-ray sources. We define the total X-ray emission from low-mass Pop III stars per galaxy as follows;
\begin{align}
& L_{\rm X, gal}=N \times L_{\rm X} ,
\label{eq18}
\end{align}
where $N$ is the number of low-mass Pop III stars in a galaxy and $L_{\rm X}$ is the typical X-ray luminosity of a single star.
We adopt $L_{\rm X}=1.0 \times 10^{28}$ erg/s as the typical luminosity for $B_{\rm c}\sim (2-6)$G that gives the peak in Figure \ref{fig4}.

For estimating $N$ in Equation (\ref{eq18}), we have to put extra assumptions about the formation of the Pop III stars, which are quite uncertain. Assuming the mass fraction $f_{\rm cor}$ of the stars with corona whose luminosity is $\approx$ $L_{\rm X}$, we obtain
\begin{align}
\begin{split}
N &= 2.5 \times 10^8 \left(\frac{\rm SFR}{1 M_{\odot} \rm yr^{-1}}\right) \left(\frac{f_{\rm cor}}{0.3}\right) \times \left(\frac{M}{0.6M_{\odot}}\right)^{-1}
\left(\frac{\tau}{0.5 \rm Gyr}\right),
\end{split}
\label{eq19}
\end{align}
where $\tau=0.5$ Gyr is the typical duration of the cosmic reionization \citep{Fa_2006}; we adopt a possible upper limit for the choice of $\tau$ as the duration of Pop III star formation throughout the reionization. In the later time of the duration of $\tau=0.5$ Gyr, second-generation low-mass Pop II stars contribute to the reionization in addition to the Pop III component.  Our previous works show that the X-ray luminosity from coronae with $Z\leq10^{-3}Z_{\odot}$ is comparable to that with $Z=0$ \citep{Su2018, WaSu2019}. Therefore, $\tau=0.5$ Gyr is a reasonable estimate. We assume that all the Pop III stars with $0.3M_{\odot} < M < 0.9M_{\odot}$ emit $L_{\rm X}=1.0\times 10^{28}$erg s$^{-1}$. We adopt the normalization, $f_{\rm cor}=0.3$, from the Kroupa IMF \citep{Kr2001, WeKr2004}.
$0.6M_{\odot}$ in Equation (\ref{eq19}) is chosen as the median stellar mass in this range.  In addition, SFR is here assumed to be $1M_{\odot}$yr$^{-1}$ per galaxy during $\tau$.
We note that SFR and the mass fraction we used are those in the local universe for simplicity because there are little constraints on the formation of low-mass Pop III stars. Therefore, $f_{\rm cor}$ should be smaller if the Pop III IMF has a top-heavy shape (e.g., \citealp{BrLa2004}), and accordingly the value of SFR would be suppressed.

We then obtain the ratio of X-ray emission to the SFR,
\begin{align}
& \frac{L_{\rm X, gal}}{\rm SFR} = 2.5 \times 10^{36} \left(\frac{L_{\rm X}}{10^{28}\rm erg \hspace{0.8mm}s^{-1}}\right)\rm{erg/s} \left( M_{\odot} \rm{yr}^{-1} \right) .
\label{eq20}
\end{align}

Referred to previous studies, the expected ratio for the HMXBs is $\sim 10^{40} f_{\rm X}\rm{erg/s} ( M_{\odot} \rm{yr}^{-1} )$ \citep{Fu2006, Fr_2013}, where $f_{\rm X}$ is the X-ray efficiency and its standard value is $f_{\rm X}=1$. This relation is based on the observation of nearby galaxies and we here assume that it can be extrapolated to higher redshifts. However, $f_{\rm X}$ in higher redshifts is still under investigation and it is sometimes considered to be higher values \citep{Me_2013, Xu_2014}. 
In addition, the average contribution from mini-quasars is about 10\% of that from XRBs, $\sim 10^{39}f_{\rm X}$ erg s$^{-1}$ \citep{Fi_2014}.  If we adopt the normalized value, $L_{\rm X}=10^{28}$erg s$^{-1}$, based on our results, the contribution from the low-mass Pop III stars is negligible in comparison to those from XRBs and mini-quasars. % and they do not give a significant impact on the cosmic reionization.

Next, we evaluate the effect on IGM heating by the soft X-rays.
We first consider the energy conservation in the expanding universe;
\begin{align}
& \frac{\mathrm{d}T_{\rm K}}{\mathrm{d}t} = -2H(z)T_{\rm K} + \sum_{i} \frac{2}{3}\frac{\epsilon_i}{k_{\rm B} n_{\rm H}},
\end{align}
where $T_{\rm K}$ is the kinetic temperature of the gas and $n_{\rm H}$ is the number density of neutral hydrogen. $H(z)$ is the Hubble parameter and $\epsilon_i$ is the energy injected to the IGM through process $i$. 
We here focus on X-ray heating by low-mass stars and describe the energy as follows;
\begin{align}
& \epsilon_X = 2.5\times 10^{36} f_{\rm heat} \dot\rho_{\rm SFR} \hspace{3mm} \mathrm{erg \hspace{1mm} cm}^{-3} s^{-1},
\end{align}
where we adopt the $L_{\rm X}$-SFR relation from Equation (\ref{eq20}). $f_{\rm heat}$ is the fraction of the X-ray energy that is used to heat the IGM. The value of $f_{\rm heat}$ depends on the ionized fraction of the gas and the X-ray photon energy and we adopt $f_{\rm heat}=0.25$, which is slightly higher than that of other X-ray sources owing to the softer spectra of low-mass stars \citep{Ric_2002, FuSt2010}.  $\dot \rho_{\rm SFR}$ is the star formation rate density and we use $\dot\rho_{\rm SFR}=10^{-2} M_{\odot} \rm{yr}^{-1}\rm{Mpc}^{-3}$ at $z=9$ from the parameterized model of the evolving $\dot \rho_{\rm SFR}$ in \cite{Ro2015}. 
Then, the contribution to the gas temperature due to X-ray heating is 
\begin{align}
\begin{split}
 \Delta T_{\rm K} &=\frac{2}{3}\frac{\epsilon_X}{k_{\rm B}n_{\rm H} H(z)} \\ &\approx  1.4 \times 10^{-1} \left(\frac{f_{\rm heat}}{0.25}\right) 
 \left(\frac{\dot\rho_{\rm SFR}}{10^{-2} M_{\odot} \rm{yr}^{-1} \rm{Mpc}^{-3}}\right)\left(\frac{1+z}{10}\right) \rm K,
\label{eq23}
\end{split}
\end{align}
which is below the cosmic microwave background temperature $T_{\rm CMB}=2.73(1+z)$ K. %Therefore, the low-mass stars cannot preheat the gas enough in the early epoch.  %while it is even smaller than $\approx 10^3 f_{\rm X}\frac{1+z}{10}$ K which is estimated with massive Pop III remnants \citep{Fu2006} so that low-mass Pop III stars do not give a significant impact on the cosmic thermal history.

Therefore, if we take our simulations at face value, the low-mass Pop III stars do not make a critical contribution to the IGM heating.
However, we mention a possibility that they produce stronger radiation which is not considered in our simulations. As we noted in Section \ref{sec41}, young solar-type stars are magnetically active and emit strong X-rays. The low-mass Pop III stars that could contribute to the reionization should be inevitably young ones so that it is important to take their property into account. While the saturated X-ray luminosity, $L_{\rm X} \approx 4\times 10^{30}$ erg/s, of solar-type stars is 400 times larger than our results $L_{\rm X}\approx 10^{28}$ erg/s, such strong emission would be also possible for Pop III stars with long loops and strong magnetic environment. 
If $L_{\rm X}$ of young Pop III stars is comparable to the saturated luminosity of solar-type stars, Equation (\ref{eq20}) then yields $L_{\rm X, gal}/\rm{SFR} \approx 1.0 \times 10^{39}$ erg/s ($M_{\odot}\rm{yr}^{-1}$).

In addition, we showed that the X-ray radiation from moderate-sized loops are stronger for smaller metallicity \citep{WaSu2019}.
Therefore, the saturated luminosity of low-mass Pop III stars is possible to exceed $L_{\rm X}\geq 4\times 10^{30}$ erg/s. 
%If this was applicable to very active stars with longer loops, the saturated luminosity of low-mass Pop III stars is inferred to excess that of the solar-type stars, $L_{\rm X}\geq 4\times 10^{30}$ erg/s. 
Accordingly, $L_{\rm X,gal}/\rm{SFR}$ would be even larger than the above value and it could be comparable to that of other X-ray objects. It also provides a larger effect on the IGM heating; the gas temperature from the X-ray heating could be more than $\Delta T_{\rm K}\approx 50$ K, which exceeds the CMB temperature at high redshifts. While the heating by other sources such as HMXBs is still more effective (e.g., $\sim 10^3 f_{\rm X} \frac{1+z}{10}$ K in \cite{Fu2006}), low-mass stars may have a role to promote the preheating due to their softer energy spectra and bring forward the moment of the heating transition at which $T_{\rm gas} = T_{\rm CMB}$ to appear in the 21cm signals. Given the strong X-ray activities which we have not considered in this study, low-mass Pop III stars could be a reasonable contributor to the cosmic reionization. To examine the coronal properties of young Pop III stars, we plan to perform the simulations with strong magnetic fields and long loops for our future work.

The low-mass Pop III stars with the simulated masses emit the EUV ($\geq 13.6$eV) radiation with $L_{\rm EUV} \sim 10^{28}$ erg/s in a wide range of $B_{\rm c}$ (Figure \ref{fig3}). 
 On the other hand, the massive metal-free stars have the effective temperature $T_{\rm eff}\sim 10^5$K which has a weak dependence on stellar mass \citep{Sc2002,Br_2001} and emit a large amount of ionizing radiation. For instance, the stars with $M_{\star}= 120M_{\odot}$ emit $1.4\times 10^{50}$ hydrogen ionizing photons per second during their lifetime $\tau_{\rm life} = 2.5$Myr \citep{Sc2002}. Assuming that all the photons have an energy of $h\nu \sim 13.6$ eV, the emission from a massive Pop III star roughly corresponds to $L_{\rm EUV} \approx 10^{39}$ erg/s which is much larger than the typical $L_{\rm EUV}$ from a low-mass star; we here note that, since the EUV radiation from low-mass stars is weighted on the higher-energy side than that from massive stars, the efficiency of the ionization is higher owing to multiple times of ionization by an ionizing photon. %Although $L_{\rm UV}$ from a massive Pop III star is much larger than $L_{\rm EUV}$ from a low-mass counterpart, in order to compare the bulk contributions the IMF has to be determined, which we discuss later in this subsection. 
 
We next estimate how much gas the emitted EUV photons from low-mass Pop III stars can ionize.  We determine the Str\"{o}mgren radius with the expected number of EUV photons $N_{\rm EUV}$ as follows;
 \begin{align}
 R_{\rm s} = \left(\frac{3N_{\rm EUV}}{4\pi \alpha}\right)^{1/3} n_e^{-2/3}, 
\end{align}
where $\alpha = 2.6 \times 10^{-13}$ cm$^3$s$^{-1}$ is the hydrogen recombination coefficient at $T=10^4$ K. Using our typical results of $L_{\rm EUV}\sim 10^{28}$ erg/s and assuming that all the photons have the same energy of $h\nu=13.6$ eV,
\begin{align}
 R_{\rm s} = 7.5\times 10^{16} \mathrm{cm} \left(\frac{N_{\rm EUV}}{4.6\times10^{38} s^{-1}}\right)^{1/3} \left(\frac{n_e}{1\rm cm^{-3}}\right)^{-2/3}.
 \label{eq24}
\end{align}
Incidentally, a similar calculation with a massive star ($N_{\rm EUV}=1.4 \times 10^{59}$ /s) gives $R_{\rm s}\approx 5.0\times 10^{20}$ cm. Because the radiation from a low-mass Pop III star is much smaller than that from a higher-mass counterpart, the ionized region produced by a single stars is also small. %However, in order to compare the bulk contributions the IMF has to be determined, which we discuss later in this subsection.  

Finally, we estimate the total contribution of low-mass Pop III stars to the cosmic reionization.
We can determine the total mass of the ionized region in a galaxy using Equation (\ref{eq19}) and (\ref{eq24}); %It is obtained from multiplying the mass of a Str\"{o}mgren sphere $\frac{4}{3}\pi R_s ^3$ by $N$;
\begin{align}
\begin{split}
 M_{\mathrm{ion}} &=  \frac{4}{3}\pi R_{\mathrm{s}}^3 n_{\mathrm{p}} m_{\mathrm{p}} N \\
  &= 3.75 \times 10^{2}  M_{\odot} \left(\frac{n_{\rm p}}{1 \rm{cm}^{-3}}\right) \left(\frac{N}{2.5\times 10^8}\right),
\end{split}
\end{align}
where $n_{\rm p}$ and $m_{\rm p}$ are the proton number density and mass.
If we consider the ionized fraction inside a Milky Way-like galaxy, the ionized mass by low-mass stars accounts for only $\sim 10^{-9}-10^{-8}$ of the total baryon mass of a galaxy $\sim (0.8-1)\times 10^{11} M_{\odot}$ \citep{Bl_2016}, while for massive stars $M_{\rm ion}=1.1\times 10^7 M_{\odot}$ which accounts for $\sim 10^{-4}$ of the total mass.

\section{Summary}
\label{sec5}

We performed the simulations for the heating of coronal loops on metal-free stars with various stellar masses and magnetic environments. 
We found that the physical condition of the corona and its radiative property are certainly affected by the coronal magnetic field strengths in relation to the nonlinearity of the Alfv\'{e}nic waves, $\delta B/B_s$.  
For weak $B_c$ cases that give large $\delta B/B_s$, most of the Alfv\'{e}nic waves dissipate in the low atmosphere and cannot heat the corona to $>10^6$K. These cases with low-temperature coronae give only weak X-ray radiation, while the moderate level of the EUV radiation is still emitted from the broad regions of the loop.

On the other hand, for strong $B_c$ cases, a large fraction of the waves can propagate upward owing to the small $\delta B/B_s$ and efficiently heat the corona.
In the regime of $B_{\rm c}<10$ G, the coronal temperature and density  increase with $B_{\rm c}$ and the hot corona is kept stable. Accordingly the loop emits larger EUV and X-rays.
%$L_{\rm EUV}$ and $L_{\rm X}$ take the maximum at $B_{\rm c}~(2-6)$G. 

However, when $B_{\rm c}>10$G, the radiative cooling is enhanced as the coronal density increases by the chromospheric evaporation, which triggers thermally unstable cyclic evolution of  the temperature. 
The time-averaged coronal temperature and density decrease with increasing $B_{\rm c}>10$ G; both $L_{\rm EUV}$ and $L_{\rm X}$ also drop. 

We note that the behavior at $B_{\rm c}>10$ G may be because of the limitation of our treatment; our simulations only consider the shock heating without incorporating other heating processes. It may change the trend at $B_{\rm c}>10$G; if the turbulent heating is considered, $L_{\rm EUV}$ and $L_{\rm X}$ would not be reduced for large $B_{\rm c}$, while the trends for weak $B_{\rm c}$ would not be affected because most of the input waves are damped in the low atmosphere.

 Based on our results, we evaluated the contribution from the low-mass Pop III coronae to ionizing and heating the IGM at high-redshift. Although they can continuously emit the radiations owing to their long lifetime, $L_{\rm EUV}$ and $L_{\rm X}$ from each star are much weaker than those of massive Pop III stars and other X-ray objects. Therefore it seems to have a small effect on the cosmic reionization. 
 %However, since the emitted energy by low-mass Pop III stars lies in a different range from other candidates of the reionization, it may contribute to ionizing the gas by working differently than others.

However, we would like to emphasize that there is still a possibility for the contribution of low-mass Pop III coronae to the reionization.
The magnetic activity level is correlated with stellar age.
The age-$L_{\rm X}$ relation of solar-type stars indicates that $L_{\rm EUV}$ and $L_{\rm X}$ from young low-mass Pop III stars can be by an order of magnitude larger than those of our results. If low-mass Pop III stars emit strong X-rays that are inferred from the saturated level of solar-type stars, their contribution is not negligible but could be comparable to that from other sources.

%\label{sec:maths} % used for referring to this section from elsewhere

\section*{Acknowledgements}
We thank Anastasia Fialkov and Naoki Yoshida for useful discussions and comments.
This work was supported by Grants-in-Aid for Scientific Research from the MEXT of Japan, 17H01105. Numerical computations were in part carried out on PC cluster at Center for Computational Astrophysics, National Astronomical Observatory of Japan.
%The Acknowledgements section is not numbered. Here you can thank helpful
%colleagues, acknowledge funding agencies, telescopes and facilities used etc.

%%%%%%%%%%%%%%%%%%%%%%%%%%%%%%%%%%%%%%%%%%%%%%%%%%
\section*{Data Availability}
The data underlying this article will be shared on reasonable request to the corresponding author.



\bibliographystyle{mnras}
\bibliography{main} % if your bibtex file is called example.bib

%\begin{thebibliography}{99}
%\end{thebibliography}

%%%%%%%%%%%%%%%%%%%%%%%%%%%%%%%%%%%%%%%%%%%%%%%%%%

%%%%%%%%%%%%%%%%% APPENDICES %%%%%%%%%%%%%%%%%%%%%

%\appendix


%%%%%%%%%%%%%%%%%%%%%%%%%%%%%%%%%%%%%%%%%%%%%%%%%%

% Don't change these lines
\bsp	% typesetting comment
\label{lastpage}
\end{document}